%% file: paper.tex
\documentclass[aps,prd,reprint,USenglish,floatfix,nofootinbib]{revtex4-1}

\pdfoutput=1

\usepackage[lineno=false,biblatex=false]{atlaspackage}

\usepackage{atlasphysics}
\hypersetup{pdftitle={ATLAS draft},pdfauthor={The ATLAS Collaboration}}

\usepackage[preprint=true]{atlascover}

\newcommand{\alpgen}{{\textsc{Alpgen}}\xspace}

\newcommand{\herwig}{{\textsc{Herwig}}\xspace}

\newcommand{\cteqsixl}{{\textsc{CTEQ6L1}}\xspace}
\newcommand{\ctten}{{\textsc{CT10}}\xspace}
\newcommand{\pythiaeight}{{\textsc{Pythia8}}\xspace}

\newcommand{\pythiasix}{{\textsc{Pythia6}}\xspace}
\newcommand{\geant}{{\textsc{GEANT4}}\xspace}

\newcommand{\powheg}{{\textsc{Powheg}}\xspace}

\newcommand{\perugia}{{\textsc{Perugia2011C}}\xspace}
\newcommand{\auet}{{\textsc{AUET2}}\xspace}

\newcommand{\au}{{\textsc{AU2}}\xspace}

\newcommand{\AtlasCoordFootnote}{%
ATLAS uses a right-handed coordinate system with its origin at the nominal interaction point (IP)
in the center of the detector and the $z$-axis along the beam pipe.
The $x$-axis points from the IP to the center of the LHC ring,
and the $y$-axis points upwards.
Cylindrical coordinates $(r,\phi)$ are used in the transverse plane, 
$\phi$ being the azimuthal angle around the $z$-axis.
The pseudorapidity is defined in terms of the polar angle $\theta$ as $\eta = -\ln \tan(\theta/2)$.
Angular distance is measured in units of $\Delta R \equiv \sqrt{(\Delta\eta)^{2} + (\Delta\phi)^{2}}$.}

\AtlasTitle{\sisetup{detect-all=true}Search for Higgs bosons decaying to $aa$ in the $\mu\mu\tau\tau$ final state in \textit{pp} collisions at $\sqrt{s} = $~\SI{8}{\tera\electronvolt} with the ATLAS experiment\sisetup{detect-all=false}}

\date{\today}

\PreprintIdNumber{CERN-PH-EP-2015-057}

\AtlasJournal{Phys. Rev. D}

\AtlasAbstract{%
  A search for the decay to a pair of new particles of either the \SI{125}{\giga\electronvolt} Higgs boson ($h$) or a second $CP$-even Higgs boson ($H$) is presented.  The dataset correspods to an integrated luminosity of \SI{20.3}{\per\femto\barn} of $pp$ collisions at $\sqrt{s}=$~\SI{8}{\tera\electronvolt} recorded by the ATLAS experiment at the LHC in 2012.  The search was done in the context of the next-to-minimal supersymmetric standard model, in which the new particles are the lightest neutral pseudoscalar Higgs bosons ($a$).  One of the two $a$ bosons is required to decay to two muons while the other is required to decay to two $\tau$-leptons. No significant excess is observed above the expected backgrounds in the dimuon invariant mass range from \SIrange{3.7}{50}{\giga\electronvolt}.  Upper limits are placed on the production of $h\rightarrow{aa}$ relative to the Standard Model $gg\rightarrow{h}$ production, assuming no coupling of the $a$ boson to quarks.  The most stringent limit is placed at \SI[number-unit-product = {}]{3.5}{\percent} for $m_a =$ \SI{3.75}{\giga\electronvolt}.  Upper limits are also placed on the production cross section of $H\rightarrow{aa}$ from \SIrange{2.33}{0.72}{\pico\barn}, for fixed $m_a$ =~\SI{5}{\giga\electronvolt} with $m_H$ ranging from \SIrange{100}{500}{\giga\electronvolt}.
}

\begin{document}

\title{\AtlasTitleText}

\author{The ATLAS Collaboration}

\begin{abstract}
	\AtlasAbstractText
\end{abstract}

\maketitle

\section{Introduction}
\label{sec:intro}

  Since its introduction, the Standard Model (SM) has successfully predicted several new particles, culminating in the discovery of the Higgs boson ($h$)  in July 2012~\cite{Aad:2012tfa,cms_higgs}.  The $h$ boson discovered by ATLAS and CMS appears to have properties consistent with that predicted by the SM within current experimental uncertainties~\cite{ATLAS-CONF-2015-007,Khachatryan:2014jba}.  While the many achievements of the SM are remarkable, several unanswered questions remain, such as the hierarchy problem~\cite{Weinberg:1975gm,Gildener:1976ai,Weinberg:1979bn,Susskind:1978ms} and the nature of the observed dark-matter abundance in the universe.  By introducing an additional symmetry between fermions and bosons, Supersymmetric (SUSY) extensions~\cite{Miyazawa01121966,PhysRevD.3.2415,Golfand:1971iw,Neveu:1971rx,Neveu:1971iv,Gervais:1971ji,Volkov:1973ix,Wess:1973kz,Wess:1974tw} of the SM are able to address both of these issues.

  The minimal supersymmetric SM (MSSM)~\cite{Fayet:1976et,Fayet:1977yc,Farrar:1978xj,Fayet:1979sa,Dimopoulos:1981zb} is the simplest extension to the SM that incorporates supersymmetry.  It predicts four additional Higgs bosons, generally assumed to be heavier than the $h$ boson: two neutral states, the $H$ and $A$ bosons, as well as two charged states, the $H^{\pm}$ charged bosons.  However, the recent discovery of only a $h$ boson and several null-result searches for the associated enhancements in Higgs boson decay rates~\cite{ATLAS_MSSM_tautau,Aad:2014kga,CMStautau} have made the model less attractive.  Specifically, the measured mass of the Higgs boson~\cite{Aad:2014aba,Khachatryan:2014jba} close to \SI{125}{\giga\electronvolt} results in the reintroduction of a small hierarchy problem in the MSSM, which is removed in the next-to-MSSM (NMSSM)~\cite{Ellwanger:2001iw,Curtin:2013fra} by allowing for additional contributions to the mass of the Higgs boson from new scalar particles.  In particular, the NMSSM contains an additional pseudoscalar Higgs boson ($a$), generally assumed to have a mass lower than the $h$ boson since its mass is protected by a Peccei--Quinn symmetry~\cite{PhysRevD.79.115006}.  The \SI[number-unit-product = {}]{95}{\percent} confidence level (CL) upper limit on the branching ratio (BR) of the $h$ boson decaying to non-SM particles is independently set by the ATLAS and CMS experiments to be \SI[number-unit-product = {}]{37}{\percent} and \SI[number-unit-product = {}]{49}{\percent} respectively~\cite{ATLAS-CONF-2014-010,Khachatryan:2014jba}.

  The analysis presented in this paper targets the $H\rightarrow{aa}$ decay for $a$ boson masses above the $\tau$-lepton pair-production threshold.  The current lower limit on $m_{H}$ for such decays was set by the ALEPH experiment at LEP to be $m_{H} >$~\SI{107}{\giga\electronvolt} for BR($H\rightarrow{aa})$~$=$~1, under the assumption that the $H$ boson couples to the $Z$ boson with SM-strength $Zh$ coupling~\cite{Schael:2010aw}. The D\O~experiment at the Tevatron also set limits on $H\rightarrow{aa}$ in this $m_{a}$ range in the $\mu\mu\tau\tau$ channel~\cite{Abazov:2009yi}.  At the LHC, the CMS experiment looked for direct $a$ boson production with $a\rightarrow\mu\mu$, with $m_a$ between \SI{5.5}{\giga\electronvolt} and \SI{14}{\giga\electronvolt}~\cite{PhysRevLett.109.121801}.  Another search for the $a$ boson below the $\tau$-lepton pair-production threshold has been performed by the CMS experiment, which looked for the 4$\mu$ final state~\cite{tagkey2013564}.

  This analysis uses data corresponding to an integrated luminosity of \SI{20.3}{\per\femto\barn} of $pp$ collisions at  $\sqrt{s}$~$=$~\SI{8}{\tera\electronvolt} recorded by the ATLAS detector at the LHC in 2012.  A search is performed for events consistent with the $H\rightarrow{aa}$ process, where one $a$ boson decays to two muons and the other decays to a pair of $\tau$-leptons with at least one $\tau$-lepton decaying to an electron or muon.  The search is performed in a range of $m_a$ values from \SIrange{3.7}{50}{\giga\electronvolt}, with $m_H$ set to \SI{125}{\giga\electronvolt} to be consistent with the measured mass of the $h$ boson, $m_h$~\cite{Aad:2014aba,Khachatryan:2014jba}.   A range of $m_H$ values from \SIrange{100}{500}{\giga\electronvolt}, with $m_a$ set to \SI{5}{\giga\electronvolt}, is also considered.  For the purposes of this analysis, the assumption is made that there is no coupling of the $a$ boson to quarks, and the event selection has been optimized for sensitivity to $m_a \lesssim~$\SI{10}{\giga\electronvolt} (approximately the $b\bar{b}$ production threshold), for which the decay products of either $a$ boson are likely to overlap.  In this range, restricting the decays of one $a$ boson to a pair of muons (rather than allowing for both to decay to pairs of $\tau$-leptons) reduces the total production rate of the signal by a factor of approximately 100.  This decrease in signal efficiency is accepted in exchange for a very high trigger efficiency, a larger signal-to-background ratio, and an expected narrow resonance in the dimuon invariant mass ($m_{\mu\mu}$) spectrum.  The latter is used to discriminate between background and signal hypotheses based on templates derived in both data and simulation. 
  
\section{ATLAS detector}
\label{sec:detector}

The ATLAS experiment~\cite{atlas-detector} at the LHC is a multi-purpose particle detector
with a forward-backward symmetric cylindrical geometry and a near $4\pi$ coverage in 
solid angle.\footnote{\AtlasCoordFootnote}
It consists of an inner tracking detector surrounded by a thin superconducting solenoid
providing a \SI{2}{\tesla} axial magnetic field, electromagnetic and hadron calorimeters, and a muon spectrometer.
The inner tracking detector covers the pseudorapidity range $|\eta| < 2.5$.
It consists of silicon pixel, silicon microstrip, and transition radiation tracking detectors.
Lead/liquid-argon (LAr) sampling calorimeters provide electromagnetic (EM) energy measurements
with high granularity.
A hadron (iron/scintillator-tile) calorimeter covers the central pseudorapidity range ($|\eta| < 1.7$).
The endcap and forward regions are instrumented with LAr calorimeters
for EM and hadronic energy measurements up to $|\eta|$~$=$~4.9.
The muon spectrometer surrounds the calorimeters and is based on
three large air-core toroid superconducting magnets with eight coils each.
Its bending power is in the range from \SIrange{2.0}{7.5}{\tesla\metre}.
It includes a system of precision tracking chambers and fast detectors for triggering.
A three-level trigger system is used to select events.
The first-level trigger is implemented in hardware and uses a subset of the detector information
to reduce the accepted rate to at most \SI{75}{\kilo\hertz}.
This is followed by two software-based trigger levels that
together reduce the accepted event rate to \SI{400}{\hertz} on average
depending on the data-taking conditions during 2012.

\section{Monte Carlo simulation}
\label{sec:simulation}

Monte Carlo simulations are used in this analysis to model the $H\rightarrow{aa}$ signal process and to roughly estimate the composition of the SM background, which is then fitted to the data, with the exception of the low-mass SM resonances ($J/\psi$, $\psi^\prime$, $\Upsilon_{1S}$, $\Upsilon_{2S}$, and $\Upsilon_{3S}$).  All Monte Carlo (MC) simulated samples are obtained with the full {ATLAS} detector simulation~\cite{:2010wqa} in which the particle propagation through the detector is modeled with \geant~\cite{Agostinelli:2002hh}.  The effect of multiple proton--proton collisions from the same or nearby beam bunch crossings (in-time and out-of-time pile-up, respectively) is incorporated into the simulation by overlaying additional minimum-bias events generated with the \pythiaeight generator~\cite{Sjostrand:2007gs} onto events with a hard scattering.  Simulated events are weighted to match the distribution of the average number of interactions per bunch crossing observed in data, but are otherwise reconstructed in the same manner as data.  The event generators, cross sections, underlying-event parameter tunes, and the parton distribution function (PDF) sets used for simulating the SM background processes are summarized in Table~\ref{tab:MCsamples}.

\begin{table*}
\caption{\small\label{tab:MCsamples}
	Simulated signal and background samples.  The background simulation is used to understand the background composition and $\mu\mu$ invariant mass distribution.
	The term ``tune'' refers to the choice of parameters used for the underlying-event generation.  Limits are set on the signal cross section, so the entry here is listed as `$\cdot\cdot\cdot$'.
}
\scalebox{0.94}{
\begin{ruledtabular}
\begin{tabular}{lllll}\toprule
Process & Generator & Cross section  & Tune   & PDF set  \\\colrule
$H\rightarrow{aa}$ & \pythiaeight~\cite{Sjostrand:2007gs} & \multicolumn{1}{c}{$\cdot\cdot\cdot$} & \au~\cite{ATLAS:2012uec} & \cteqsixl~\cite{Pumplin:2002vw} \\
\colrule
$t\bar{t}$ & \powheg~\cite{Nason:2004rx,Frixione:2007vw} + \pythiaeight    & NNLO+NNLL~\cite{Cacciari:2011hy,Baernreuther:2012ws,Czakon:2012zr,Czakon:2012pz,Czakon:2013goa,Czakon:2011xx} & \perugia~\cite{Skands:2010ak} & \ctten~\cite{CT10pdf} \\ 
$Z/\gamma^{*}$+jets & \alpgen~\cite{Mangano:2002ea}+\pythiasix~\cite{Sjostrand:2006za} & NLO & \perugia & \cteqsixl \\
$b\bar{b}$ & \pythiaeight & LO & \au & \cteqsixl \\
$tW$ & \powheg+\pythiaeight   & NNLO+NNLL~\cite{Kidonakis:2010ux} & \perugia & \ctten \\
$WW$, $WZ$ & \herwig~\cite{Herwig} & NLO & \auet~\cite{ATLAS:2011zja} & \cteqsixl \\
$ZZ$ & \powheg+\pythiaeight & NLO & \au & \ctten \\
$h\rightarrow{ZZ}$ & \powheg+\pythiaeight & NNLO+NNLL~\cite{CERNYellowReport3}  & \au & \ctten \\
\end{tabular}
\end{ruledtabular}
}
\end{table*}

 In each simulated $H\rightarrow{aa}$ signal event, a scalar $H$ boson is produced via gluon fusion and is then forced to always decay to two pseudoscalar $a$ bosons.  Both the scalar $H$ and pseudoscalar $a$ bosons are expected to have narrow widths, which are negligible when compared to the detector resolution.  The $H$ width is set to the value for a SM Higgs boson with the same mass and the $a$ boson width is set to \SI{1}{\mega\electronvolt}~\cite{PhysRevLett.109.121801}.  This analysis searches for a resonance in the $m_{\mu\mu}$ spectrum and is therefore insensitive to the resolution for $m_H$.  The $a$ bosons are each allowed to decay only to either two muons or two $\tau$-leptons.  The decay rate of $a\rightarrow\mu\mu$ is expected to depend on the mass of the $a$ boson as well as the muon and $\tau$-lepton masses~\cite{PhysRevD.79.115006}:
 \begin{eqnarray} \frac{\Gamma\left(a\rightarrow\mu\mu\right)}{\Gamma\left(a\rightarrow\tau\tau\right)} = \frac{m_\mu^2}{m_\tau^2\sqrt{1-(2 m_\tau/m_a)^2}}.\label{eq:aBR}
 \end{eqnarray}
The widths values calculated from this equation are used by Pythia to generate signal events, under the assumption that BR$(a\rightarrow\mu\mu)$ = 1 - BR$(a\rightarrow\tau\tau)$.  BR$(a\rightarrow\mu\mu)$ ranges from \SI[number-unit-product = {}]{1.25}{\percent} for $m_a$ = \SI{3.7}{\giga\electronvolt} down to \SI[number-unit-product = {}]{0.35}{\percent} for $m_a$ = \SI{50}{\giga\electronvolt}.

\section{Event selection}
\label{sec:eventselection}

The data used for this analysis were collected using a combination of a single-muon trigger and an asymmetric dimuon trigger.  The transverse momentum (\pt) threshold for the single-muon trigger is \SI{36}{\giga\electronvolt}, while the dimuon trigger thresholds for the highest \pt (leading) and the second-highest \pt (subleading) muons are \SI{18}{\giga\electronvolt} and \SI{8}{\giga\electronvolt}, respectively~\cite{Aad:2014sca}.  Muons are reconstructed by combining tracks in the inner detector (ID) with those in the muon spectrometer (MS)~\cite{Aad:2014rra} and are selected if they satisfy the following criteria: each muon must have a \pt~$>$~\SI{16}{\giga\electronvolt}, $|\eta| < 2.5$, projected longitudinal impact parameter $|z_{0}\sin{\theta}|<$~\SI{0.4}{\milli\meter}, the ratio of the transverse impact parameter $d_0$ to its estimated uncertainty $\sigma_{d_0}$ such that $|d_0|/\sigma_{d_0}<3$, and a track isolation requirement.  The transverse and longitudinal impact parameters are defined with respect to the primary vertex, defined as the vertex with the largest $\sum\pt^2$ of the associated tracks.  The track isolation of each muon is defined as the scalar sum of the \pt of all tracks with \pt~$>$~\SI{1}{\giga\electronvolt}, in a surrounding $\Delta{R}$~$=$~0.3 cone, excluding those tracks associated with either muon.  The track isolation is required to be less than \SI[number-unit-product = {}]{12}{\percent} of the muon \pt.  The selected muons are organized into oppositely charged pairs, which are then ordered by the vector sum of the \pt of the constituent muons.  The leading pair is identified as the $a\rightarrow\mu\mu$ candidate.  Events are required to have an $a\rightarrow\mu\mu$ candidate with \pt~$>$~\SI{40}{\giga\electronvolt}.

\subsection{Signal regions}

Two signal regions (SR) are designed to select $H\rightarrow{aa}$ events where one $a$ boson directly decays to a pair of muons, and the other to a pair of $\tau$-leptons.  The regions are optimized for the scenario where one $\tau$-lepton decays to either a muon or an electron and the other $\tau$-lepton is identified by selecting one to three additional tracks.\footnote{The signal region has been designed to select events in which the charged decay product(s) of the second $\tau$-lepton is either an electron, a muon, a single charged hadron or are three charged hadrons; each decay product is identified as a selected track.  For the signals targeted by this analysis, one or more of the tracks in the last case can have $p_T$ insufficient to be selected.  This results in final states with less than three reconstructed tracks.}

   Each signal event must have an $a\rightarrow\mu\mu$ candidate with a dimuon invariant mass between \SI{2.8}{\giga\electronvolt} and \SI{70}{\giga\electronvolt}.\footnote{The mass range used to define the two signal regions (\SIrange{2.8}{70}{\giga\electronvolt}), and ultimately used to fit the observed $m_{\mu\mu}$ spectra, is wider than the range of signal hypotheses (\SIrange{3.7}{50}{GeV}) in order to be able to increase the stability of the fit when considering the low-mass SM resonances and the low-end tail of the SM $Z$ resonance.}  Additionally, the event must contain a third lepton (muon or electron), with the flavor indicated in the SR name (i.e. SR$\mu$ or SR$e$).  In the case where the third lepton is a reconstructed muon, it must have \pt~$>$~\SI{7}{\giga\electronvolt} and $|\eta| < 2.5$.  If the third lepton is a reconstructed electron, it is required to satisfy identification requirements optimized for good efficiency extending to low \pt~\cite{ATLAS-CONF-2014-032}, must have \pt~$>$~\SI{7}{\giga\electronvolt} and $|\eta| < 2.47$.  Electrons in the transition region between the barrel and endcap EM calorimeters ($1.37 < |\eta| < 1.55$) are rejected.  The two $a$ bosons are expected to be produced back-to-back in the transverse plane.  This topology is enhanced by requiring the azimuthal separation between the tagged $a\rightarrow\mu\mu$ candidate and the third lepton to be greater than \SI{1}{rad}.   Furthermore, the two $\tau$-leptons from the decay of the $a$ boson tend to be highly collimated for low values of $m_a$.  Therefore, the third lepton (muon or electron) is required to have one, two, or three tracks, in a cone of $\Delta{R}$~$=$~0.4, not including the track of the lepton itself.  Tracks are reconstructed in the ID and are required to have \pt~$>$~\SI{1}{\giga\electronvolt}, $|\eta| < 2.5$, transverse and longitudinal impact parameters, $|d_0|$ and $|z_0|$, less than \SI{1}{mm}, at least seven hits in the two silicon tracking detectors, and one hit in the innermost layer of the pixel detector if such a hit is expected.  The leading \pt track within the $\Delta{R}$~$=$~0.4 cone around the third lepton, but not matched to the lepton itself, is denoted the lead-track.  It is used to approximate the axis of the second $\tau$-lepton from the $a\rightarrow{\tau\tau}$ decay.  The $\Delta{R}$~$=$~0.2 cone, centered on this axis,  may have no more than three tracks, not including the track of the third lepton.  The charge of the lead-track is taken to be the same as its $\tau$-lepton ancestor and is required to be opposite to that of the lepton.  Finally, the third lepton is required to pass a cut on its track isolation, defined as the scalar sum of the \pt of the tracks that are within a cone of $\Delta{R}$~$=$~0.4.  The lead-track is excluded from this calculation. The track isolation is required to be less than \SI[number-unit-product = {}]{12}{\percent} of the lepton's \pt.
    
  If the event has more than one candidate to be the third lepton, the leading muon is used to define the signal region.  If no third muon is identified, the leading electron is used.  Events with muon as well as electron candidates are classified as SR$\mu$ events in order to maximize the signal-to-background ratio.  No veto is applied to events with more than three leptons in order to maintain signal selection efficiency for events where both $\tau$-leptons decay to muons or electrons.  In these cases, the additional lepton may be selected as the lead-track.  Table~\ref{t:sigCutFlow} shows the selection efficiency for a signal simulated with $m_a$~$=$~\SI{5}{\giga\electronvolt} and $m_H$~$=$~$m_h$~$=$~\SI{125}{\giga\electronvolt}.

\begin{table}[!htb]
\caption[]{\small\label{t:sigCutFlow}
Relative efficiency for each selection step for a signal simulated with $m_a$~ $=$ ~\SI{5}{\giga\electronvolt} and $m_H$ ~$=$ ~$m_h$ ~$=$ ~\SI{125}{\giga\electronvolt}.  The first entry is a generator-level selection relative to the number of generated $h\rightarrow{aa}\rightarrow{\mu\mu\tau\tau}$ events.\footnote{The generator-level selection requires at least two muons with \pt above \SI{15}{\giga\electronvolt} and $|\eta|<2.5$.  Generator-level events are further required to have a pair of such muons that have opposite charge and have \pt$\left(\mu\mu\right)>$~\SI{30}{\giga\electronvolt}.  Only \SI[number-unit-product = {}]{0.5}{\percent} of generated events have a $\mu\mu\tau\tau$ final state due to the small rate of $a\rightarrow\mu\mu$.  As discussed in Section~\ref{sec:models}, a small number of events with four $\tau$-leptons are also accepted, but the rate is too small to impact the efficiency reported in this table.}  The top section of the table shows the efficiency for selecting $a\rightarrow\mu\mu$ candidates while the bottom part is for the selection of events in the two signal regions.}
\scalebox{0.94}{
\begin{ruledtabular}
\begin{tabular}{ccc}
Selection & \multicolumn{2}{c}{Relative efficiency $\left(\si[number-unit-product = {}]{\percent}\right)$} \\\colrule
Generator-level & \multicolumn{2}{c}{\num{82.7+-0.0}} \\
Pass trigger & \multicolumn{2}{c}{\num{67.6+-0.3}} \\
Two selected muons & \multicolumn{2}{c}{\num{77.8+-0.3}} \\
Opposite charge $\left(\mu,\mu\right)$ & \multicolumn{2}{c}{\num{100.0+-0.0}} \\
\pt$\left(\mu\mu\right)>$~\SI{40}{\giga\electronvolt} & \multicolumn{2}{c}{\num{98.1+-0.1}} \\
\SI{2.8}{\giga\electronvolt}~$< m_{\mu\mu}<$~\SI{70}{\giga\electronvolt} & \multicolumn{2}{c}{\num{100.0+-0.0}}\\\colrule
& SR$\mu$ $\left(\si[number-unit-product = {}]{\percent}\right)$ & SR$e$ $\left(\si[number-unit-product = {}]{\percent}\right)$ \\\colrule
Third lepton & \num{18.2+-0.3} & \num{7.8+-0.2} \\
$\Delta{\phi}\left(\mu\mu,\ell\right)$ & \num{95.5+-0.4} & \num{93.7+-0.7} \\
1,2 or 3 nearby tracks & \num{91.4+-0.5} & \num{82.8+-1.1} \\
Opposite charge $\left(\ell,\mbox{lead-track}\right)$ & \num{91.2+-0.9} & \num{88.1+-1.1} \\
Lepton isolation & \num{75.5+-0.9} & \num{84.6+-1.3} \\
\end{tabular}
\end{ruledtabular}
}
\end{table}

\subsection{Validation and control regions}
\label{sec:validcontrol}

Two validation regions (VR) are used to test the methods of the analysis in a signal-free environment.  The background in the signal region is expected to be dominated by events with jets that pass the $a\rightarrow\tau\tau$ selection outlined above.  Furthermore, the charges of the reconstructed lepton and lead track in such events are not correlated. Events in the validation regions are therefore selected with the same criteria as signal events, with the sole exception that the lepton and lead-track are required to have the same charge.  The signal yield in the validation regions is expected to be less than \SI[number-unit-product = {}]{5}{\percent} (\SI[number-unit-product = {}]{10}{\percent}) of that in the muon (electron) signal region, as determined in simulation.  As done on the signal region, the two regions are denoted VR$\mu$ and VR$e$ based on the flavor of the third lepton.

Two control regions (CR) are used in this analysis to constrain the SM background, one light-flavor-dominated region CRj and one heavy-flavor-dominated region CRb.  Events in each control region are required to have an ${a\rightarrow\mu\mu}$ candidate with a mass between \SI{2.8}{\giga\electronvolt} (\SI{15}{\giga\electronvolt}) and \SI{70}{\giga\electronvolt} for CRj (CRb).\footnote{The \SI{15}{GeV} lower $m_{\mu\mu}$ limit for CRb is chosen such that the region is dominated by the $t\bar{t}$ background.}  The two regions are further defined by imposing requirements on jets in the event.  The light-flavor-dominated region is used to measure the nonresonant Drell--Yan background as well as the low-mass SM resonances.  It is defined by requiring at least one selected jet and exactly zero $b$-jets.  The heavy-flavor-dominated region must have at least two $b$-jets and is used to measure the $t\bar{t}$ background.  All events in the signal regions are explicitly removed from the control regions.

Jets are reconstructed using the anti-$k_t$ algorithm~\cite{Cacciari:2008gp} and radius parameter $R$~$=$~0.4 with three-dimensional energy clusters, measured in the calorimeter~\cite{Lampl:2008zz}, as input. 
The calibration of the clusters is performed by applying different weights to the energy deposits arising from the electromagnetic and hadronic components of the showers.  The final calibration of the jet energy corrects the response of the calorimeter to match the particle-level jet energy~\cite{Aad:2011he,Aad:2014bia}.  Corrections are first determined in simulation and then improved and validated using data.  Additional corrections for pile-up from other $pp$ interactions in the same or neighboring bunch crossings are applied as well~\cite{TheATLAScollaboration:2013pia}.
Jets are required to have \pt~$>$~\SI{40}{\giga\electronvolt} and $|\eta| < 2.5$.  Jets with \pt between \SI{40}{\giga\electronvolt} and \SI{50}{\giga\electronvolt} are further required to pass a cut on the jet vertex fraction (JVF), which is defined as the \pt-weighted fraction of tracks inside the jet that are determined to originate from the primary vertex of the event.  The JVF is required to be greater than $0.25$, but is not applied to jets that have no tracks.  Jets can be tagged as containing a $b$-hadron using a multivariate $b$-tagging algorithm, and then denoted $b$-jets.  The algorithm is based on the impact parameters of tracks and information from reconstructed secondary vertices~\cite{btag}.  
The operating point chosen for the $b$-tagging algorithm corresponds to a true $b$-jet efficiency of \SI[number-unit-product = {}]{60}{\percent} and purity of \SI[number-unit-product = {}]{95}{\percent}, as determined in $t\bar{t}$ simulation.
 
\section{Signal and background models}
\label{sec:models}

An unbinned, log-likelihood fit is performed on the observed dimuon invariant mass spectra in the signal regions (SR$\mu$ and SR$e$) to a combination of background and signal models, the details of which are described in this section.

\subsection{Signal model}

\begin{figure}[tb]
\subfloat{\includegraphics[width=0.82\columnwidth]{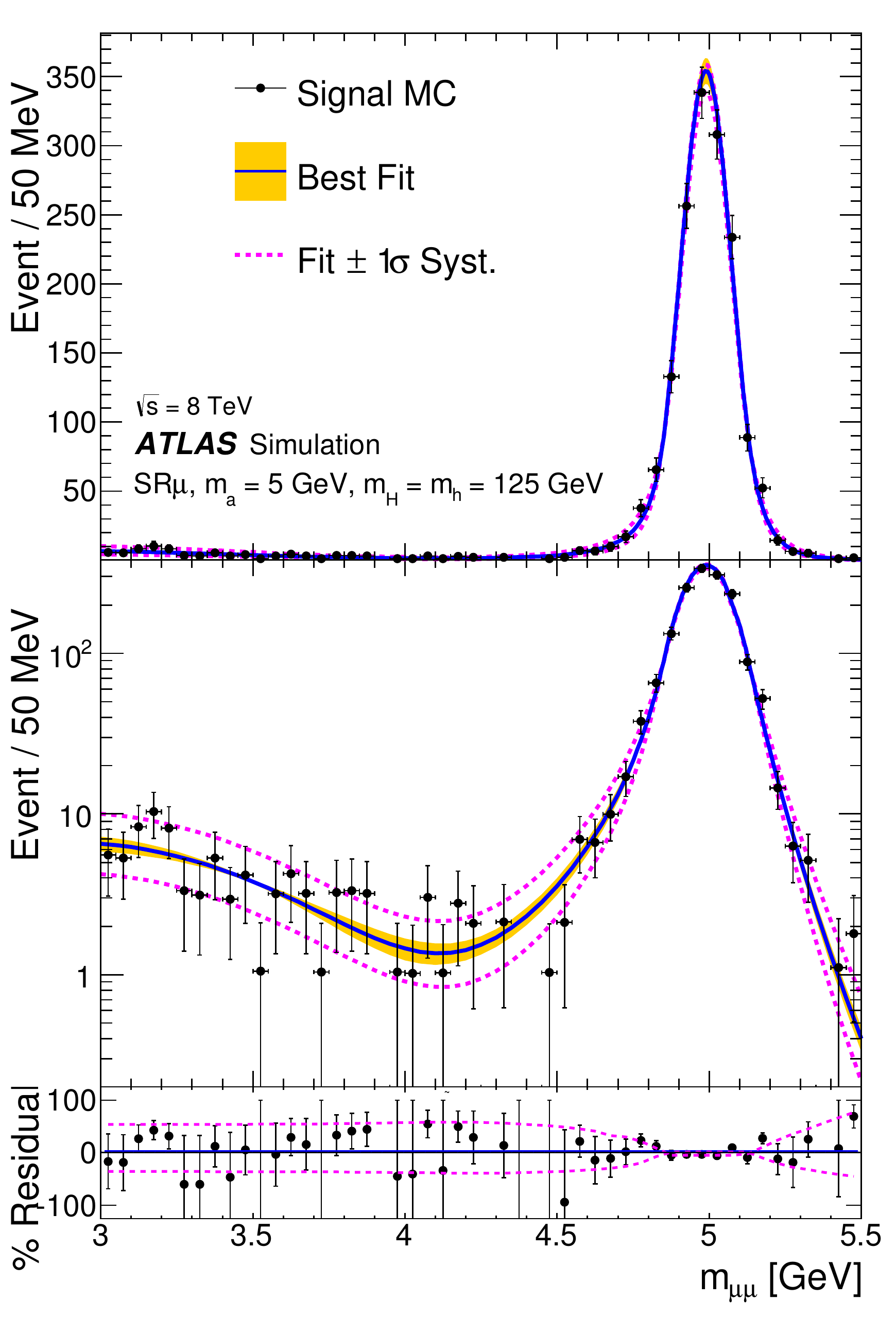}}
\caption[]{\small\label{sec:signalmodel}
Simulated dimuon invariant mass ($m_{\mu\mu}$) distribution and the result of the simultaneous fit projected into SR$\mu$ for one benchmark mass point with $m_a =$ \SI{5}{\giga\electronvolt} and $m_H=m_h=$~ \SI{125}{\giga\electronvolt}.  The best fit of the $\mu\mu$ resonance model to the $h\rightarrow{aa}$ signal simulation in SR$\mu$ is shown in blue with its uncertainty as a yellow band.  Also shown are the forced $\pm 1\sigma$ systematic uncertainty variations, defined in Sec.~\ref{sec:systematics}, in $\alpha_{\mathrm{CB}}$ and $f_{\tau\tau}$ (dashed magenta).  The top plot shows the simulation and fits on a linear scale, the middle on a logarithmic scale and the \% residual of each fit is shown at the bottom.  The errors on the signal simulation are statistical only.}
\label{fig:fitsr1}
\end{figure}

  A double-sided Crystal Ball (CB) function~\cite{Oreglia:1980cs,ATLAS-CONF-2014-031} is used for the signal model; it was found to be the simplest function that was robust enough to describe the shape of the simulated signal resonances and the observed SM resonances.\footnote{For a given resonance $X$, the double-sided Crystal Ball function has been simplified to remove those parameters that are common to all resonances.  The expression $\mbox{CB}\left(\mu_X,\sigma_X,\alpha_{L}=\alpha_{\mathrm{CB}},\alpha_{H}=\alpha_{\mathrm{CB}},n_{L}=2.5,n_{H}=10\right)$ has been shortened to $\mbox{CB}\left(\mu_X,\sigma_X\right)$.
  \begin{align}
  & \mbox{CB}\left( m_{\mu\mu} | \mu,\sigma,\alpha_{L},\alpha_{H},n_{L},n_{H} \right) = \label{eq:CB}\\
 & 
\begin{cases}
   e^{-\frac{\left(m_{\mu\mu}-\mu\right)^2}{2\sigma^2}}, & \text{for} -\alpha_{L} < \frac{m_{\mu\mu}-\mu}{\sigma} < \alpha_H \\
  A_{L} \cdot\left(B_{L}-\frac{m_{\mu\mu}-\mu}{\sigma}\right)^{-n_{L}}, & \text{for} \frac{m_{\mu\mu}-\mu}{\sigma} \leq -\alpha_{L} \\
  A_{H} \cdot\left(B_{H}-\frac{m_{\mu\mu}-\mu}{\sigma}\right)^{-n_{H}}, & \text{for} \frac{m_{\mu\mu}-\mu}{\sigma} \geq \alpha_H
\end{cases}\nonumber
\end{align}
\begin{align}
A_{i} = \left(\frac{n_{i}}{|\alpha_{i}|}\right)^{n_{i}} \cdot e^{-\frac{\alpha_{i}^2}{2}}  & & B_{i} = \frac{n_{i}}{|\alpha_{i}|}-|\alpha_{i}| \nonumber
\end{align}
}  The Crystal Ball function is composed of a Gaussian (GA) core, with mean $\mu_{\mathrm{CB}}$ and width $\sigma_{\mathrm{CB}}$, and power-law distributions of orders 2.5 and 10 for the low-end and high-end tails.  The threshold parameter $\alpha_{\mathrm{CB}}$, which is in units of $\sigma_{\mathrm{CB}}$, determines the point of transition from the core to either tail.  The mean of the Crystal Ball function is assumed to be proportional to $m_a$, with slope $a_\mu$, while its width is assumed to be linearly dependent on both $m_a$ and $m_H$ with slopes $a_\sigma$ and $b_\sigma$ respectively.

  The Crystal Ball function is used to model the lineshape of the $a\rightarrow\mu\mu$ resonance.  A small contribution from $a\rightarrow\tau\tau\rightarrow\mu\mu+4\nu$ is included as a Gaussian distribution, which, due to the kinematics of the $\tau\rightarrow\mu+2\nu$ decay, is expected to have a lower mean and worse resolution than the $a\rightarrow\mu\mu$ resonance.  The mean, $\mu_{\tau\tau}$, and width, $\sigma_{\tau\tau}$, are set proportional to the corresponding parameters of the Crystal Ball function with the parameters $k_{\mu}<1$ and $k_\sigma>1$.  The fraction of $a\rightarrow\tau\tau\rightarrow\mu\mu+4\nu$ in the total signal is given by $f_{\tau\tau}$.  The full signal model is:  
\begin{eqnarray}
P_{a} = \mbox{CB}\left(m_{\mu\mu}\mid\mu_{\mathrm{CB}},\sigma_{\mathrm{CB}}\right) + f_{\tau\tau}\mbox{GA}\left(m_{\mu\mu}\mid\mu_{\tau\tau},\sigma_{\tau\tau}\right),\nonumber\\
\mu_{\mathrm{CB}}=a_{\mu}\cdot m_{a} ~,~ \mu_{\tau\tau}=k_{\mu}\cdot\mu_{\mathrm{CB}} ~,~ \sigma_{\tau\tau}=k_{\sigma}\cdot a_{\sigma}\cdot m_{a}~, \nonumber\\
\sigma_{\mathrm{CB}}=a_{\sigma}\cdot m_{a} + b_{\sigma}\cdot(m_H - \SI{100}{\giga\electronvolt}).~~~~~~~~ \label{eq:sigmodel}
\end{eqnarray}

The values of the parameters $b_{\sigma}$, $k_{\mu}$, $k_{\sigma}$, and $f_{\tau\tau}$ are determined by fitting the signal simulation, while the parameters $\alpha_{\mathrm{CB}}$, $a_{\mu}$, and $a_{\sigma}$ are measured in data following the procedure described in the Sec.~\ref{sec:backgroundmodel}, and are found to be consistent with simulation.   
Figure~\ref{fig:fitsr1} shows the result of a simultaneous fit of both signal regions to all simulated signal samples, projected into SR$\mu$ for one benchmark mass point with $m_a =$ \SI{5}{\giga\electronvolt} and $m_H=m_h=$~ \SI{125}{\giga\electronvolt}.

\subsection{Background model}
\label{sec:backgroundmodel}

  The full background model consists of several pieces: six SM resonances ($J/\psi, \psi^\prime, \Upsilon_{1S}, \Upsilon_{2S}, \Upsilon_{3S}, Z$), a $t\bar{t}$ component, and one piece for the nonresonant continuum background (dominated by low-mass Drell--Yan events).  Each SM resonance is modeled by the same double-sided Crystal Ball function used for the signal $a\rightarrow\mu\mu$ resonance (see Sec.~\ref{sec:signalmodel}), with the offset $b_{\sigma}$ set to zero.  The mean, $\mu_{X}$, and width, $\sigma_{X}$, of each resonance $X$ are assumed to be linearly dependent on its mass ($m_{X}$) with the same slopes as in the signal model.  The measured value of the mass of each resonance is found to be consistent with the PDG best-fit value~\cite{1674-1137-38-9-090001} and are therefore constrained to the PDG value and its associated uncertainty.  The low-mass resonances are combined into two composite models, $\psi$ and $\Upsilon$, by adding the resonances of the higher spin states with fractions $f_{\psi^\prime}$, $f_{\Upsilon_{2S}}$, and $f_{\Upsilon_{3S}}$, as shown in Eqs.~\ref{eq:psi} and~\ref{eq:ups}.
  
    The background from $t\bar{t}$ production is modeled with a Rayleigh distribution~\cite{papoulis_probability}, defined by multiplying $m_{\mu\mu}$ by a Gaussian distribution with mean set to zero and width, $\sigma_{t\bar{t}}$.   The mostly Drell--Yan continuum background is modeled by an exponential decay function, with parameter $\alpha_{\gamma^*}<0$, multiplied by $m_{\mu\mu}$ raised to the power ${n_{\gamma^*}}$.  The full expression for the continuum background is $m_{\mu\mu}^{n_{\gamma*}}e^{\alpha_{\gamma*}m_{\mu\mu}}.$

  A contribution to the background from $b\bar{b}$ production followed by two semileptonic decays of $b$-hadrons to muons was also considered and found to be small in the signal region, with a $m_{\mu\mu}$ shape similar to the Drell--Yan component but with a rate about \SI[number-unit-product = {}]{1}{\percent} as large. This is expected since dimuon events from $b\bar{b}$ are highly suppressed after applying the muon isolation requirements, the muon \pt cuts, and the dimuon \pt cut. Events from double semileptonic $b$-hadron decays ($b\rightarrow{c}\mu + X \rightarrow\mu\mu + X$) are found to contribute as well, but only for $m_{\mu\mu}<$~\SI{3.5}{\giga\electronvolt} (which is below the signal region) at a rate of about \SI[number-unit-product = {}]{10}{\percent} that of Drell--Yan events.

Lastly, the full model in each region (CRj, CRb, SR$\mu$ or SR$e$) is defined by adding four background models: one each for the $\psi$ and $\Upsilon$ resonances, one for the $t\bar{t}$ component, and one for the Drell--Yan component.  The shape parameters of each component, the fractional contribution of higher spin states, and the relative contribution of $Z$ boson to the total Drell--Yan production ($f_{Z}$) are constrained by simultaneously fitting the control regions as described in Sec.~\ref{sec:bkgmeas}.  These parameters are therefore assumed to be the same in the control and signal regions.  The relative contributions of each of the four background components (quantified by the independent parameters $f_{t\bar{t}}$, $f_{\Upsilon}$, and $f_{\mathrm{Res}}$) are expected to vary between the regions and are therefore measured in the fit to the signal regions (SR$\mu$ and SR$e$).  The complete background model is shown in Eq.~(\ref{eq:bkgmodel}):

\begin{widetext}
\begin{eqnarray}
P_{\psi} = &\mbox{CB}\left(m_{\mu\mu}\mid\mu_{J/\psi},\sigma_{J/\psi}\right) + f_{\psi^\prime}\mbox{CB}\left(m_{\mu\mu}\mid\mu_{\psi^\prime},\sigma_{\psi^\prime}\right),\label{eq:psi}\\
P_{\Upsilon} = \mbox{CB}\left(m_{\mu\mu}\mid\mu_{\Upsilon_{1S}},\sigma_{\Upsilon_{1S}}\right) + f_{\Upsilon_{2S}}\left(\mbox{CB}\left(m_{\mu\mu}\mid\mu_{\Upsilon_{2S}},\sigma_{\Upsilon_{2S}}\right) + f_{\Upsilon_{3S}}\mbox{CB}\left(m_{\mu\mu}\mid\mu_{\Upsilon_{3S}},\sigma_{\Upsilon_{3S}}\right)\right),\label{eq:ups}\\
P_{b} = m_{\mu\mu}^{n_{\gamma^*}}e^{\alpha_{\gamma^*}m_{\mu\mu}} +  f_{Z}\mbox{CB}\left(m_{\mu\mu}\mid\mu_{Z},\sigma_{Z}\right) +  f_{t\bar{t}}\mbox{R}\left(m_{\mu\mu}\mid\sigma_{t\bar{t}}\right) + f_{\mathrm{Res}}\left(P_{\psi} + f_{\Upsilon}P_{\Upsilon}\right), \label{eq:bkgmodel}\\
\mu_{X}=a_{\mu}\cdot m_{X} ~;~ \sigma_{X}=a_{\sigma}\cdot m_{X} ~;~ X = \{ J/\psi, \psi^\prime, \Upsilon_{1S}, \Upsilon_{2S}, \Upsilon_{3S}, Z\}.\nonumber
\end{eqnarray}
\end{widetext}

\section{Background measurement}
\label{sec:bkgmeas}

\begin{figure}[!htb]
\subfloat{\includegraphics[width=0.82\columnwidth]{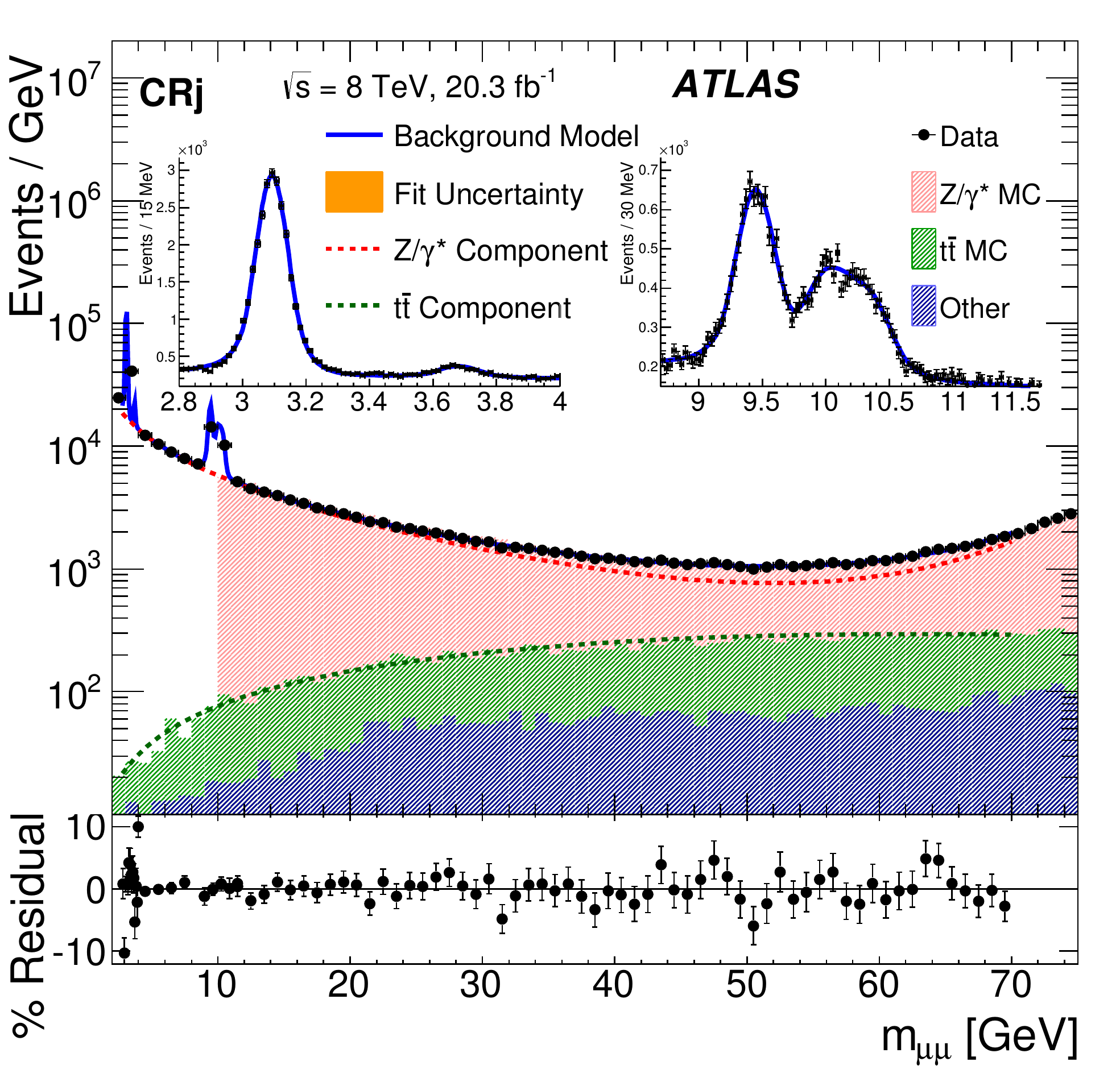}}\\
\subfloat{\includegraphics[width=0.82\columnwidth]{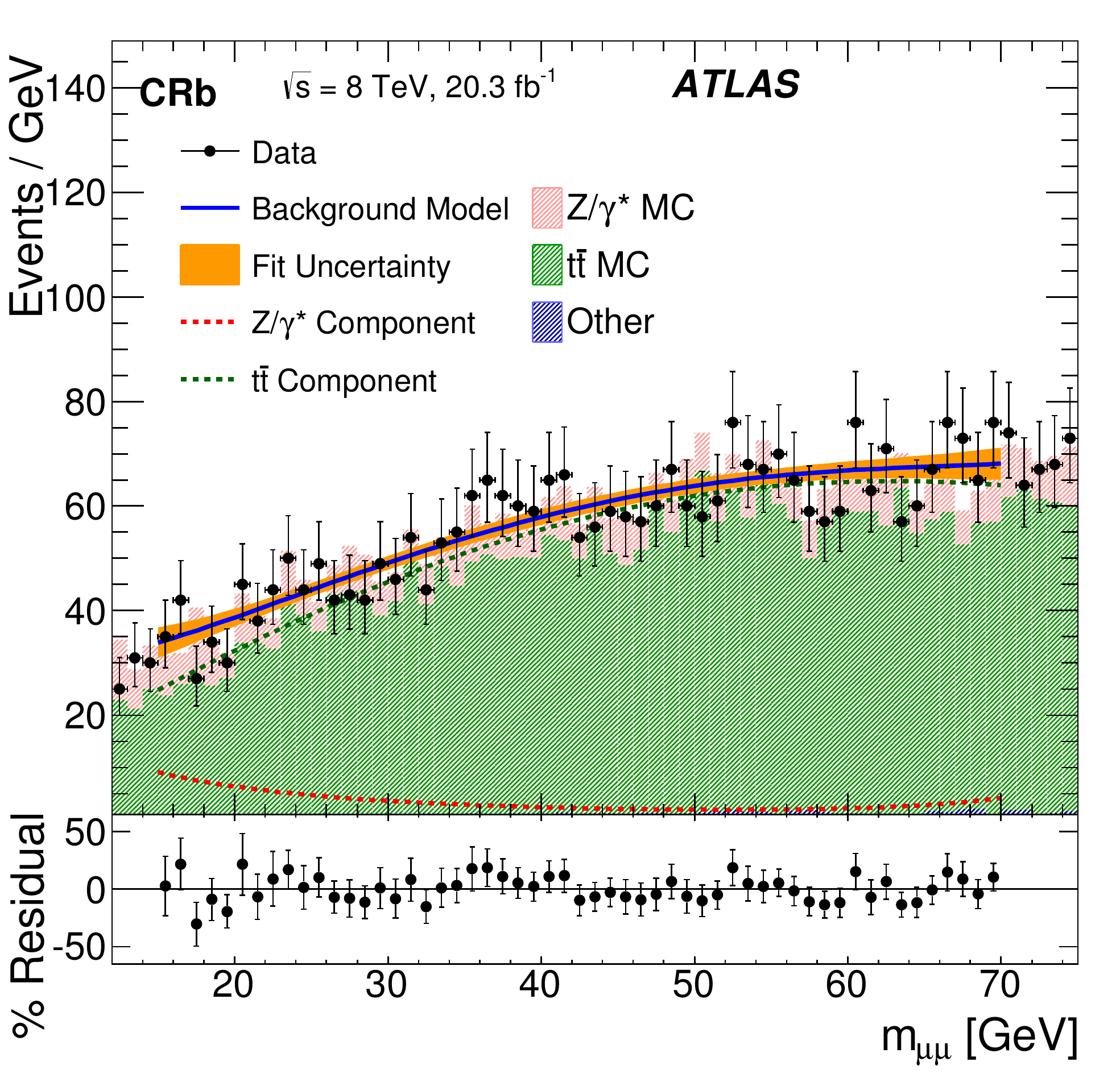}}
\caption[]{\small\label{fig:crfit}
Observed $m_{\mu\mu}$ distribution in CRj (top) and CRb (bottom) and the SM background model after a simultaneous fit.  The $Z/\gamma^*$ component of the fit is the combination of the $Z$ boson resonance and the $\gamma^*$ continuum models.  The \si[number-unit-product = {}]{\percent} residual of the fit is shown below each plot.  Simulated SM backgrounds are shown in the stack, with the $Z/\gamma^{*}$ sample only valid above $m_{\mu\mu}>$\SI{10}{\giga\electronvolt}.  The two insets show magnified versions of the $J/\psi$ and $\Upsilon$ resonances.}
\end{figure}

A simultaneous fit of the full background model is performed on the $m_{\mu\mu}$ data in the two control regions (CRj and CRb), with $m_{\mu\mu}$ in the range from \SI{2.8}{\giga\electronvolt} (\SI{15}{\giga\electronvolt}) to \SI{70}{\giga\electronvolt}, for CRj (CRb).  The results of the fit are used to constrain the dependent parameters of the four background components as described in Sec.~\ref{sec:backgroundmodel}.  The fitted values and uncertainties of the relevant parameters are reported in Table~\ref{t:crfit} and the fitted $m_{\mu\mu}$ spectra in the two regions are shown in Fig.~\ref{fig:crfit}.  A large correlation is found between $a_{\mu}$, $a_{\sigma}$, and $\alpha_{CB}$ while a large anti-correlation is found between $\sigma_{t\bar{t}}$ and $f_{Z}$.  The figure also includes a comparison with Drell--Yan and $t\bar{t}$ simulation normalized to the data.\footnote{The Drell--Yan simulation has only been produced with $m_{\mu\mu}$ above \SI{10}{\giga\electronvolt}, so is only valid for comparing to the data in that region.}  In both control regions, the data are well described by the background model, as indicated by the small residuals shown in the figure and by the good agreement with simulation.  The large number of events in the control regions provides a tight constraint on the parameters of the model.  The small shift of the resonance mean masses to \SI[number-unit-product = {}]{99.86+-0.01}{\percent} of their PDG values~\cite{1674-1137-38-9-090001}, is consistent with the $O(\SI[number-unit-product = {}]{0.1}{\percent})$ systematic uncertainty on the muon energy scale.  The fractional mass resolution, $a_\sigma$, is found to be \SI[number-unit-product = {}]{1.68+-0.02}{\percent}, corresponding to an \SI{84+-1}{\mega\electronvolt} width for a \SI{5}{\giga\electronvolt} dimuon resonance.

\begin{table}[!htb]
\caption[]{\small\label{t:crfit}
Measured values and uncertainties for the SM background model's parameters, which are constrained by a simultaneous fit to the control regions.  The errors on the fit parameters are statistical only.}
\scalebox{0.94}{
\begin{ruledtabular}
\begin{tabular}{lr|lr}
Par. & Value & Par. & Value \\\colrule
$a_{\mu}$ & \SI[number-unit-product = {}]{99.86+-0.01}{\percent} & $\sigma_{t\bar{t}}$ & \SI{60.7+-3.8}{\giga\electronvolt} \\
$a_{\sigma}$ & \SI[number-unit-product = {}]{1.68+-0.02}{\percent} & $f_{Z}\left[\frac{Z}{\gamma^*+Z}\right]$ & \SI[number-unit-product = {}]{23.4+-0.5}{\percent} \\
$\alpha_{\mathrm{CB}}$ & \num{1.49+-0.03} & $f_{\psi^\prime}\left[\frac{2S}{1S+2S}\right]$ & \SI[number-unit-product = {}]{6.3+-0.3}{\percent} \\
$\alpha_{\gamma^*}$ & \SI{-31+-3}{\per\tera\electronvolt} & $f_{\Upsilon_{3S}}\left[\frac{3S}{2S+3S}\right]$ & \SI[number-unit-product = {}]{46.8+-1.4}{\percent}\\
$n_{\gamma^*}$ & \num{-0.75+-0.02} & $f_{\Upsilon_{2S}}\left[\frac{2S+3S}{1S+2S+3S}\right]$ & \SI[number-unit-product = {}]{49.9+-0.6}{\percent}\\
\end{tabular}
\end{ruledtabular}
}
\end{table}

\section{Systematic uncertainties}
\label{sec:systematics}

\subsection{Signal model}

  In the signal model for the $a\rightarrow\mu\mu$ resonance, the parameters $\alpha_{\mathrm{CB}}$, $a_{\mu}$, and $a_{\sigma}$ are limited by the experimental resolution, and are thus assumed to be \SI[number-unit-product = {}]{100}{\percent} correlated with the corresponding parameters of the fits to the SM resonances and are measured in the observed CRj data.  To account for the extrapolation uncertainty from parameters for the SM resonances to those for signal, an additional uncertainty is assigned to the Crystal Ball function's threshold parameter, $\alpha_{\mathrm{CB}}$.  Due to the large correlation between the three parameters, this additional uncertainty assigned to $\alpha_{\mathrm{CB}}$ covers the uncertainty of the other two. The uncertainty is determined by fitting the signal model in each simulated signal sample separately and taking the maximum difference between them.
  Another assumption in the signal model is that the parameter, $b_{\sigma}$, and the Gaussian description of the $a\rightarrow\tau\tau\rightarrow\mu\mu+4\nu$ tail, parameterized by  $k_{\mu}$, $k_{\sigma}$, and $f_{\tau\tau}$, are properly described in the simulation.  The systematic uncertainty on $\alpha_{\mathrm{CB}}$ covers the uncertainty on $b_{\sigma}$, since it is strongly correlated with $\alpha_{\mathrm{CB}}$, $a_{\mu}$, and $a_{\sigma}$.  The other three parameters are similarly correlated with one another and, thus, only one additional systematic uncertainty is introduced.   The final results of the analysis are found to have little sensitivity to the amount of $a\rightarrow\tau\tau\rightarrow\mu\mu$, so a conservative \SI[number-unit-product = {}]{50}{\percent} systematic uncertainty is assigned to $f_{\tau\tau}$.

\subsection{Signal normalization}

  The dominant source of systematic uncertainty on the signal normalization is found to be the theoretical uncertainty on the rate of SM $gg\rightarrow{h/H}$ production.  In the $m_H$ range relevant for this analysis (from \SIrange{100}{500}{\giga\electronvolt}) the total uncertainty varies from \SIrange[number-unit-product = {}]{10}{11}{\percent} and is determined from the spread of the cross-section predictions using different PDF sets and their associated uncertainties, as well as from varying the factorization and renormalization scales~\cite{CERNYellowReport3}.  A constant \SI[number-unit-product = {}]{11}{\percent} is used in this analysis.
    The next largest systematic uncertainty is on the \pt resolution of the lead-track and is found to be \SI[number-unit-product = {}]{5}{\percent}.  This uncertainty on the signal normalization is determined by varying the \pt of each track by a conservative $\pm$\SI[number-unit-product = {}]{2}{\percent}, and propagating the effect through the full analysis.
      Additional sources of systematic uncertainty include those on the trigger efficiency, the lepton reconstruction efficiency, the lepton energy scale and resolution, and the charge of the track.  All of these sources were found to contribute a negligible amount to the total uncertainty on the normalization of the signal.

\subsection{Background model}

  The results of the background measurement reported in Sec.~\ref{sec:bkgmeas} are fitted values and associated uncertainties of the dependent parameters of the four background components---Drell--Yan, $t\bar{t}$, $\psi$, and $\Upsilon$.  In the fit to the signal regions, each parameter is constrained by a Gaussian prior with a mean equal to its fitted value from the background measurement and  width equal to the corresponding uncertainty.
  
  Two general assumptions are made in the construction of the background model.   First, the chosen functional form accurately describes the background in the signal regions.  Second, the dependent parameters are the same in each region.  Both of these assumptions introduce a potential bias in the final result, whereby a nonzero signal may be observed incorrectly.  The tolerance of the background model for such a spurious signal is measured for values of $m_{a}$ and $m_{H}$ corresponding to each simulated signal point.  The measurement is performed using a large sample of signal-free events.  Simulated $t\bar{t}$ events with an identified $a\rightarrow\mu\mu$ candidate are used as a large sample of events for the $t\bar{t}$ background component.  In lieu of simulation, the observed data in the light-flavor-dominated control region are used for the Drell--Yan component.  The simulated $t\bar{t}$ events are then weighted and combined with the data such that the relative contribution of $t\bar{t}$ matches the simulation-based expectation in the signal region with $m_{\mu\mu}$ between \SI{20}{\giga\electronvolt} and \SI{60}{\giga\electronvolt}.\footnote{For the spurious signal calculation, a narrower $m_{\mu\mu}$ range (\SIrange{20}{60}{\giga\electronvolt} instead of \SIrange{2.8}{70}{\giga\electronvolt}) is used to scale the $t\bar{t}$ simulation because it is the range in which the $t\bar{t}$ background is expected to dominate.}  Finally, the resulting sample is scaled to the expected normalization of the signal region.
  The combined signal and background model is fit to the large sample of events.  The potential bias is taken to be the measured rate of spurious signal + 1$\sigma$ and is found to be between \SI[number-unit-product = {}]{0.2}{\percent} and \SI[number-unit-product = {}]{3.2}{\percent} of the signal rate normalized to the SM $gg\rightarrow{h}$ production rate with BR$(h\rightarrow{aa})$ = \SI[number-unit-product = {}]{100}{\percent}, with a maximum at $m_{a}=$~\SI{20}{\giga\electronvolt} and $m_H=m_h=$~\SI{125}{\giga\electronvolt}.  The measured bias is taken as an additional uncertainty on the signal normalization.

\begin{figure}
\subfloat{\includegraphics[width=0.82\columnwidth]{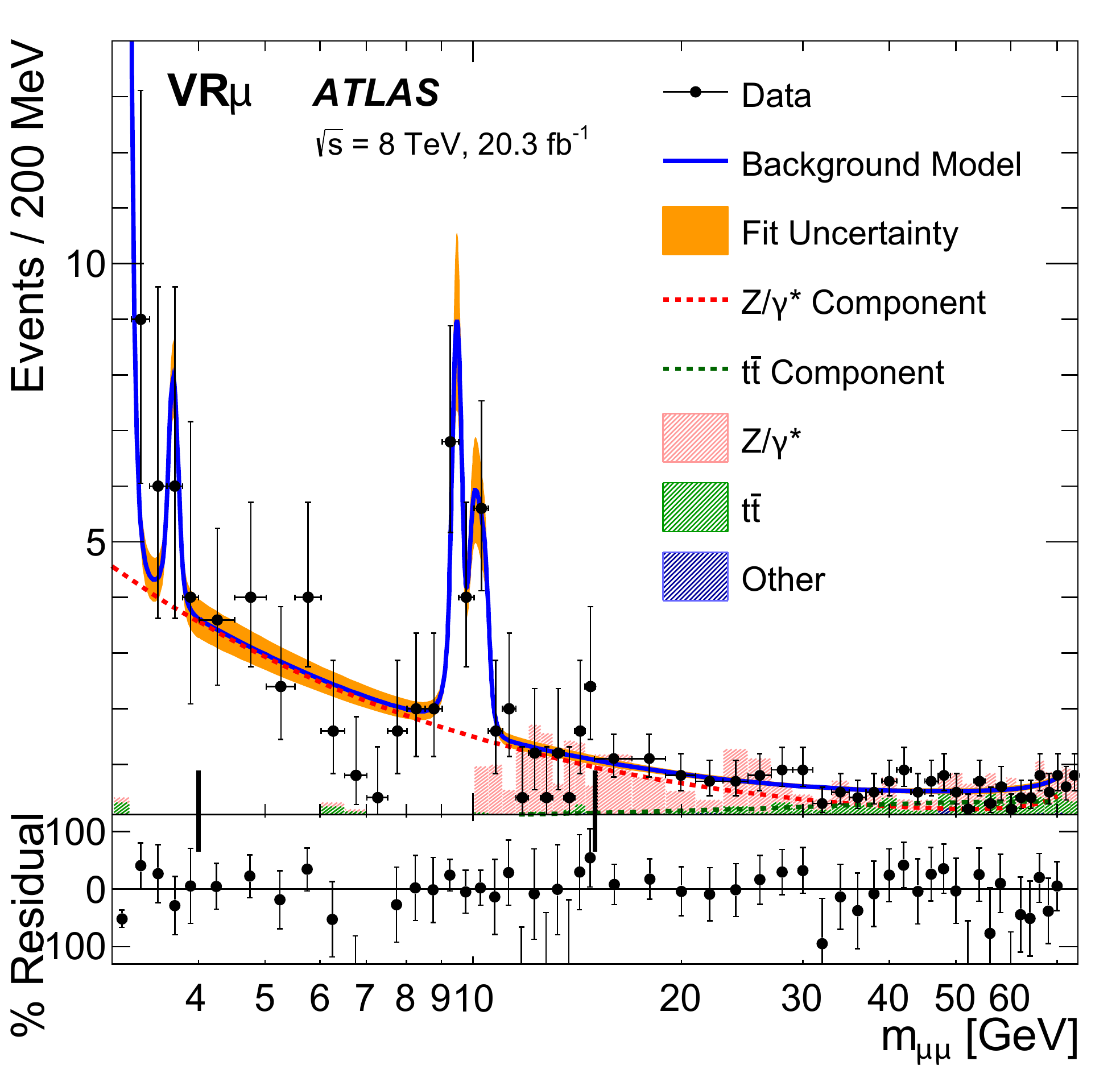}}\\
\subfloat{\includegraphics[width=0.82\columnwidth]{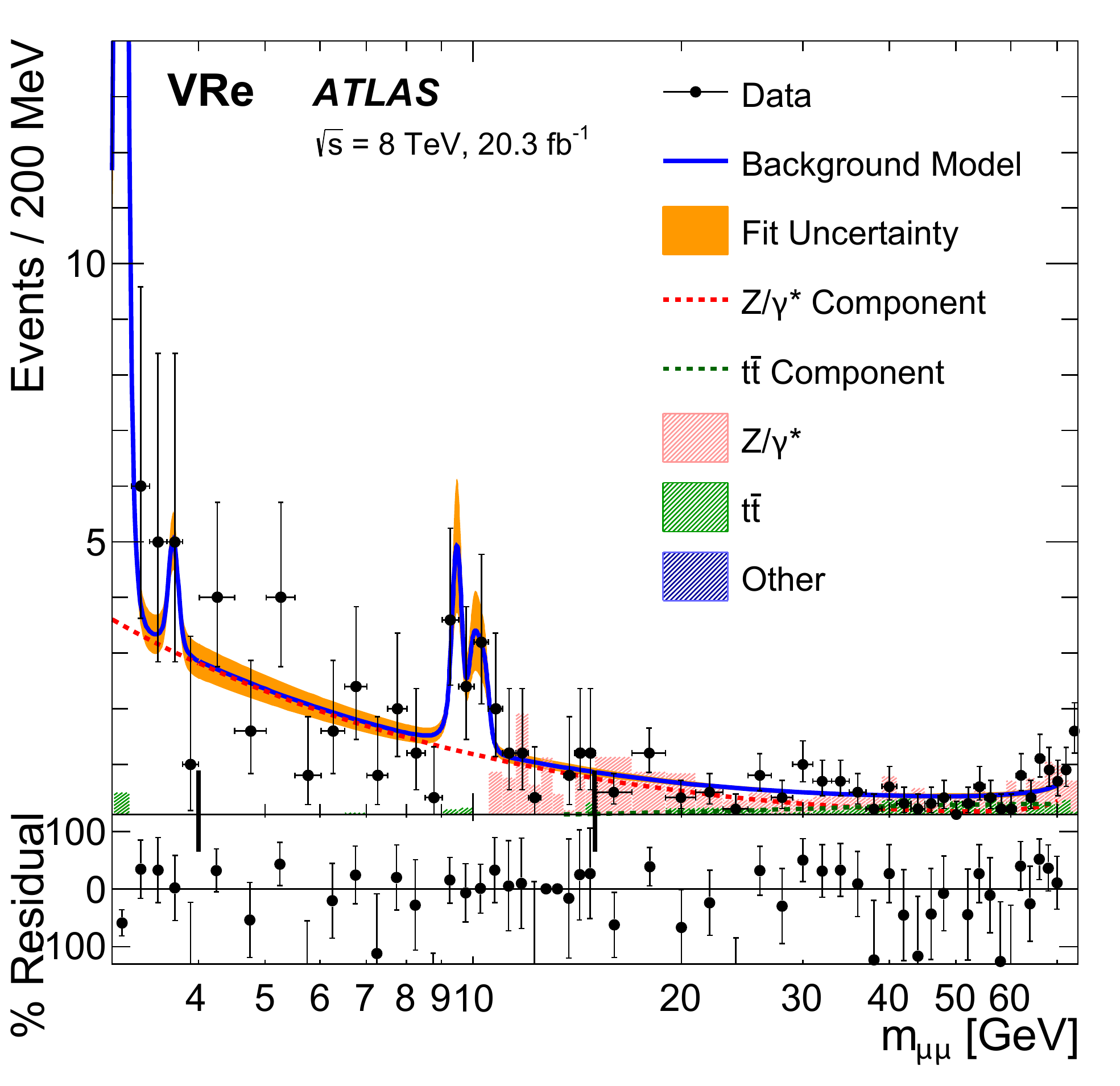}}
\caption[]{\small\label{fig:vrfit}
Observed $m_{\mu\mu}$ distribution in VR$\mu$ (top) and VR$e$ (bottom) and the background-only fit.  The $Z/\gamma^*$ component of the fit is the combination of the $Z$ boson resonance and the $\gamma^*$ continuum models.  The \si[number-unit-product = {}]{\percent} residuals are shown below each plot.  Bins below \SI{4}{\giga\electronvolt} are \SI{200}{\mega\electronvolt} wide, between \SI{4}{\giga\electronvolt} and \SI{15}{\giga\electronvolt} they are \SI{500}{\mega\electronvolt} wide, and above \SI{15}{\giga\electronvolt} they are \SI{2}{\giga\electronvolt} wide.  Simulated SM backgrounds are shown in the stack, with the $Z/\gamma^{*}$ sample only valid above $m_{\mu\mu}>$ \SI{10}{\giga\electronvolt}.}
\end{figure}

\section{Validation of methods}

  To test the methods of the analysis, two validation regions, VR$\mu$ and VR$e$ (as defined in Sec.~\ref{sec:validcontrol}), are used in place of SR$\mu$ and SR$e$.  The validation regions are designed to have properties similar to the signal regions and to also test the robustness of the method against variations in the background composition, since no \textit{a priori} assumptions are made about the relative contributions from $t\bar{t}$, Drell--Yan, $J/\psi$ or $\Upsilon$.  Furthermore, the validation checks for non-negligible backgrounds with dimuon invariant mass distributions that are unaccounted for in the background model.   The results of the simultaneous fit of the full background model (including all relevant systematic uncertainties) to the validation regions are shown in Fig.~\ref{fig:vrfit}.  In the fit, all region-independent parameters are constrained by the results of the fit to the control regions (reported in Table~\ref{t:crfit}).  The strong correlations between some of the region-independent parameters, which are reported in Sec.~\ref{sec:bkgmeas}, are not explicitly accounted for in the fit to the two validation regions.  This simplification is found to have a negligible effect on the results of the fit.  The consistency with the background-only model is evaluated by scanning the local $p$-value as a function of $m_{\mu\mu}$ from \SIrange{3.7}{50}{\giga\electronvolt}, calculated using frequentist hypothesis tests based on the profile-log-likelihood ratio test statistic and approximated with the asymptotic formulae~\cite{Cowan:2010js}.  The $p$-values are evaluated in \SI{50}{\mega\electronvolt} intervals below \SI{15}{\giga\electronvolt}, then \SI{100}{\mega\electronvolt} intervals up to \SI{30}{\giga\electronvolt}, and \SI{200}{\mega\electronvolt} intervals up to $m_{\mu\mu}=$~\SI{50}{\giga\electronvolt}.  The minimum local $p$-value is found for $m_{\mu\mu} =$~\SI{47.4}{\giga\electronvolt} to be 0.0074, corresponding to a local significance of 2.44$\sigma$.  Correcting for the look-elsewhere effect~\cite{lle} gives a global $p$-value of $0.31$, indicating that at least one excess of this magnitude, or larger, is expected from background fluctuations in at least \SI[number-unit-product = {}]{31}{\percent} of experiments.

\begin{figure}[tb]
\subfloat{\includegraphics[width=0.82\columnwidth]{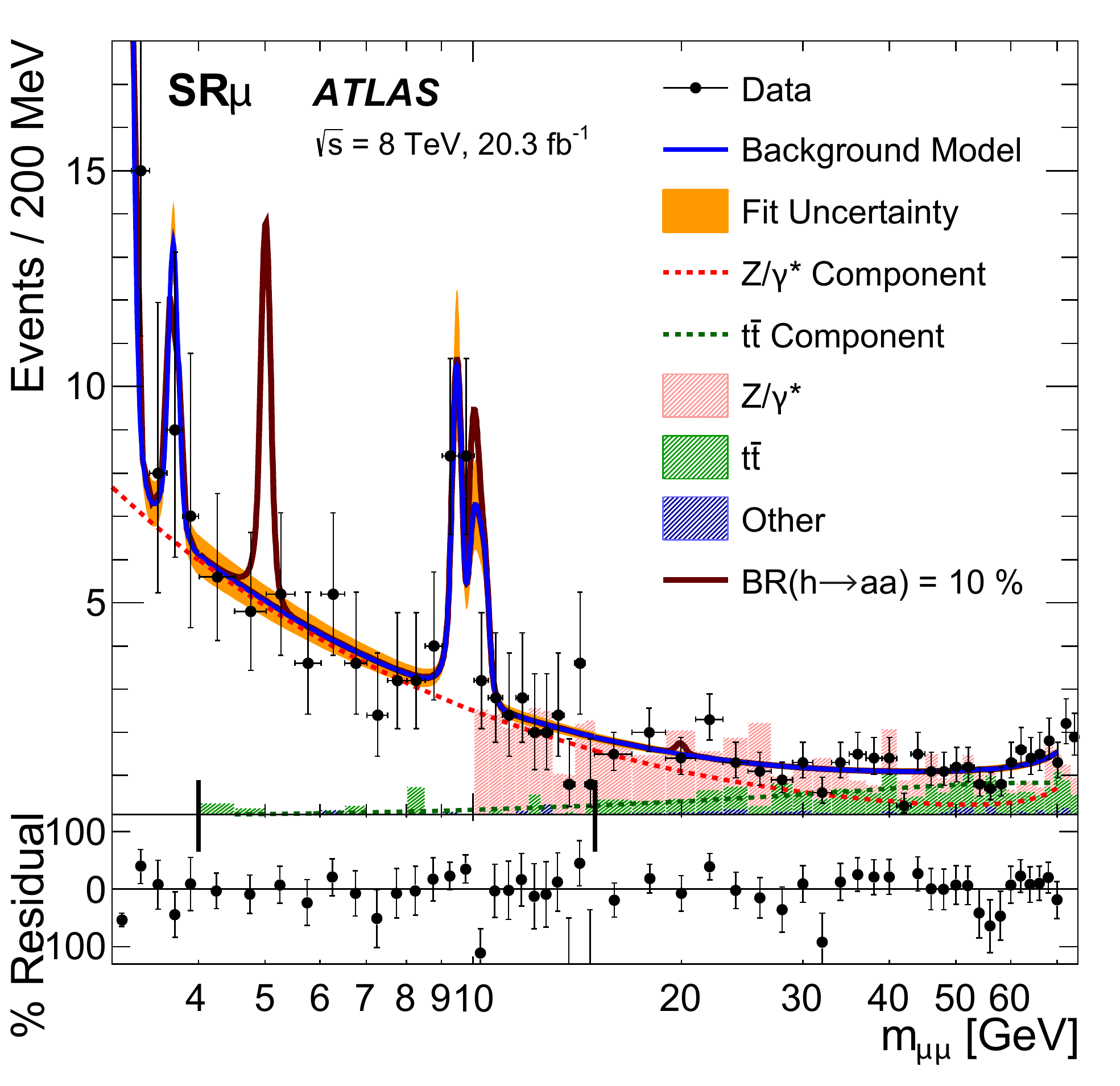}}\
\subfloat{\includegraphics[width=0.82\columnwidth]{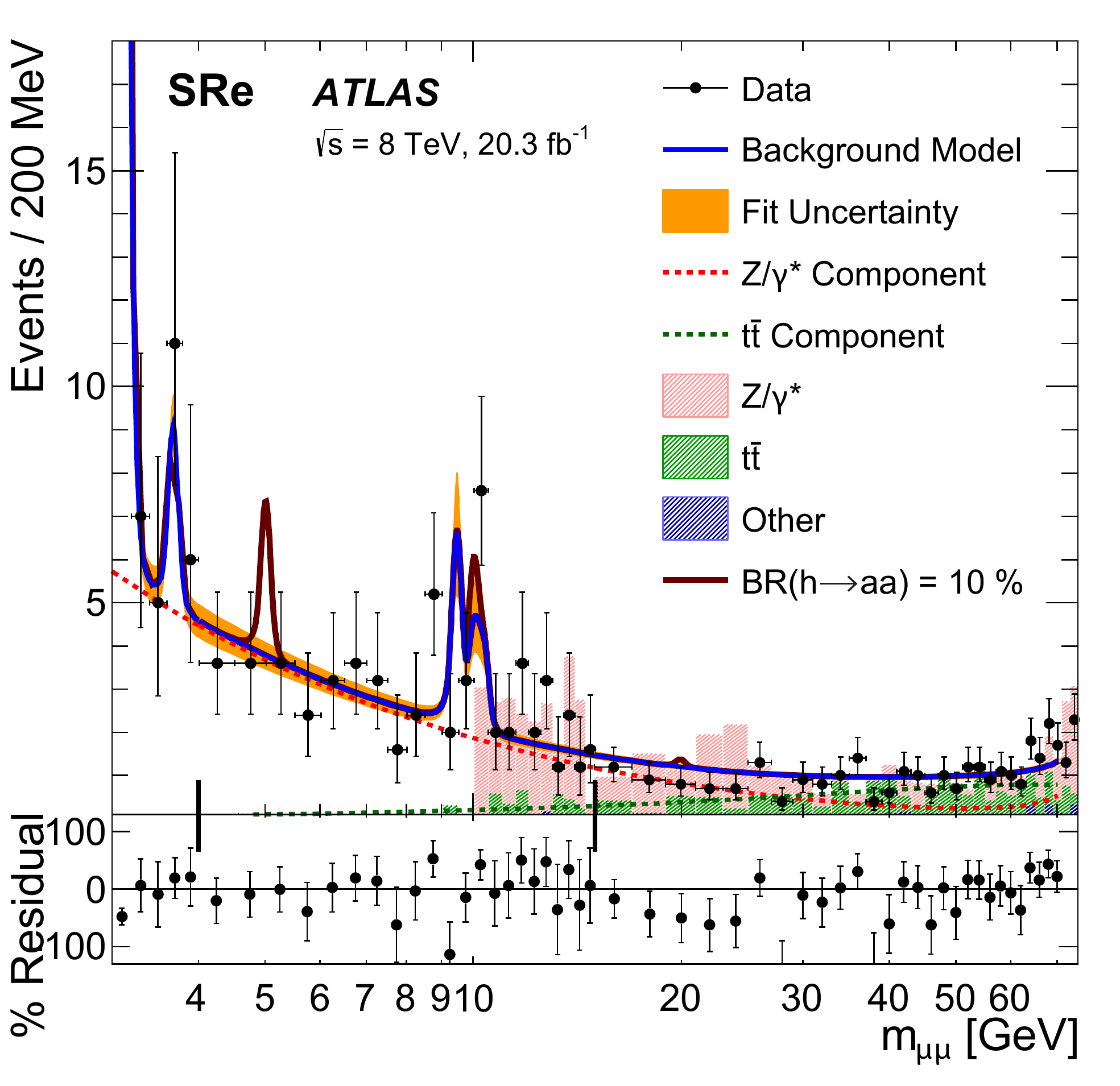}}
\caption[]{\small\label{f:MmumuSRunblind}
Observed $m_{\mu\mu}$ distribution in SR$\mu$ (top) and SR$e$ (bottom) and the background-only fit.  The $Z/\gamma^*$ component of the fit is the combination of the $Z$ boson resonance and the $\gamma^*$ continuum models.  The \si[number-unit-product = {}]{\percent} residuals are shown below each plot.  Bins below \SI{4}{\giga\electronvolt} are \SI{200}{\mega\electronvolt} wide, between \SI{4}{\giga\electronvolt} and \SI{15}{\giga\electronvolt} they are \SI{500}{\mega\electronvolt} wide, and above \SI{15}{\giga\electronvolt} they are \SI{2}{\giga\electronvolt} wide.  The expected distribution from a signal with BR$(h\rightarrow{aa})$=\SI[number-unit-product = {}]{10}{\percent} is shown for three different $m_a$ hypotheses (\SI{5}{\giga\electronvolt}, \SI{10}{\giga\electronvolt}, and \SI{20}{\giga\electronvolt}).  Simulated SM backgrounds are shown in the stack, with the $Z/\gamma^{*}$ sample only valid above $m_{\mu\mu}>$ \SI{10}{\giga\electronvolt}.}
\end{figure}

\section{Results}
\label{sec:result}

\begin{figure}[tb]
\subfloat{\includegraphics[width=0.82\columnwidth]{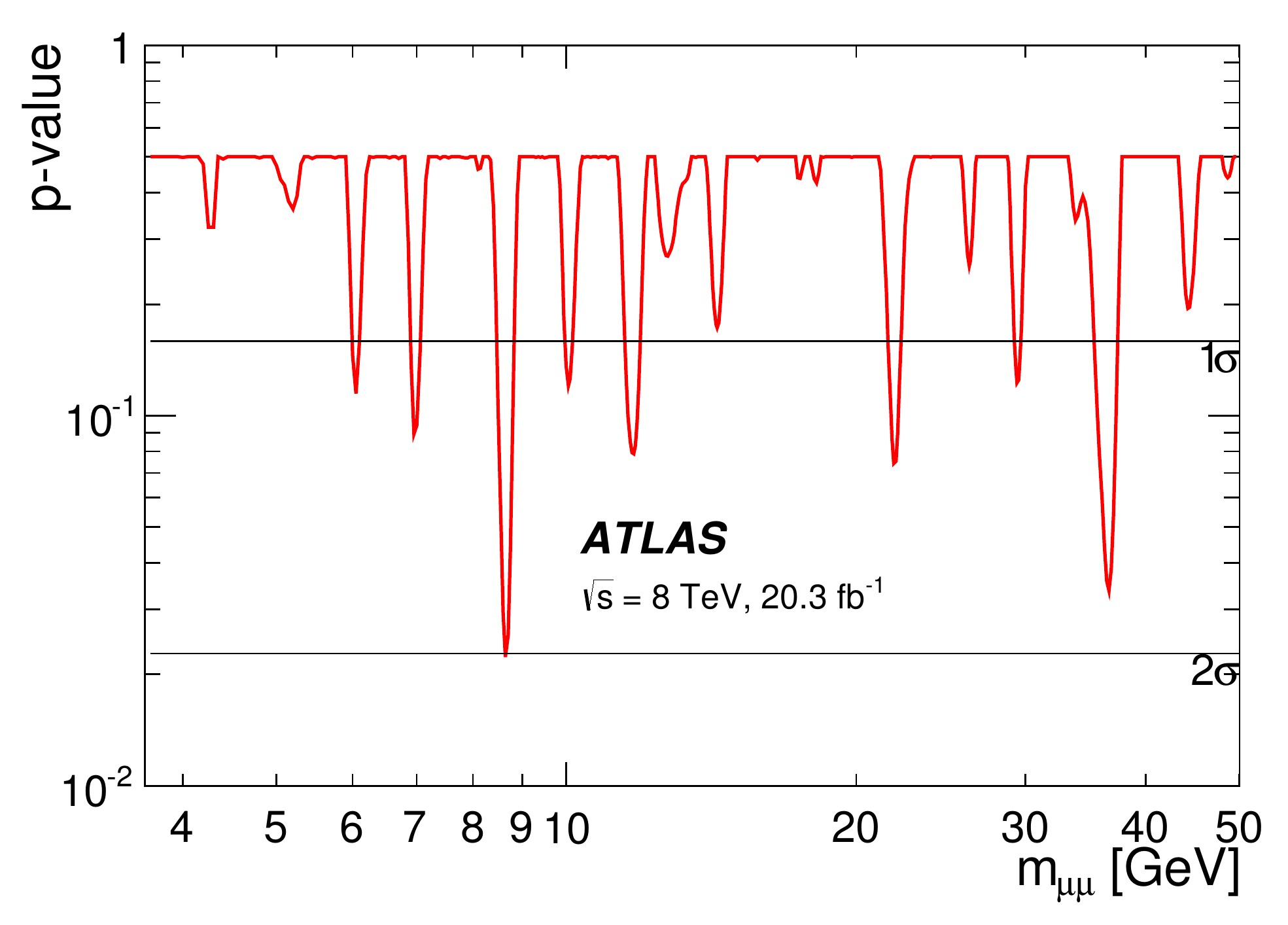}}
\caption[]{\small\label{f:pval}
Observed $p$-value as a function of $m_{\mu\mu}$, with downward fluctuations of the data represented by a $p$-value of 0.5.  The $p$-values are evaluated in \SI{50}{\mega\electronvolt} intervals below \SI{15}{\giga\electronvolt}, then \SI{100}{\mega\electronvolt} intervals up to \SI{30}{\giga\electronvolt}, and \SI{200}{\mega\electronvolt} intervals up to $m_{\mu\mu}=$~\SI{50}{\giga\electronvolt}.  The $p$-values shown have not been corrected for the look-elsewhere effect.}
\end{figure}

   A simultaneous fit of the full background model is performed on the $m_{\mu\mu}$ spectra in the two signal regions (SR$\mu$ and SR$e$), with $m_{\mu\mu}$ in the range from \SI{2.8}{\giga\electronvolt} to \SI{70}{\giga\electronvolt}.  The observed $m_{\mu\mu}$ distribution and the background-only fit are shown in Fig.~\ref{f:MmumuSRunblind}; the data are well described by the fit.  In the fit, all region-independent parameters are constrained by the results of the fit to the control regions, reported in Table~\ref{t:crfit}.  The strong correlations between some of the region-independent parameters, which are reported in Sec.~\ref{sec:bkgmeas}, are not explicitly accounted for in the fit to the two signal regions.  This simplification is found to have a negligible effect on the results of the fit.  The fitted values and uncertainties of the remaining parameters, as well as the corresponding values from the fits to the control and validation regions, are shown in Table~\ref{t:fit}.  No significant correlations are found between the parameters listed in Table~\ref{t:fit}.  The consistency with the background-only model is evaluated by scanning the local $p$-value as a function of $m_{\mu\mu}$ from \SIrange{3.7}{50}{\giga\electronvolt}, using the same calculation, $m_a$ range, and intervals used in the scan of the validation region. The results of this scan are reported in Fig.~\ref{f:pval}.  The minimum local $p$-value is found for $m_{\mu\mu} =$~\SI{8.65}{\giga\electronvolt} to be 0.0223, corresponding to a local significance of 2.01$\sigma$.  Correcting for the look-elsewhere effect~\cite{lle} gives a global $p$-value $>0.5$, indicating that at least one excess of this magnitude, or larger, is expected from background fluctuations in at least \SI[number-unit-product = {}]{50}{\percent} of experiments.

\begin{table}[!htb]
\caption[]{\small\label{t:fit}
Measured values and uncertainties of region-dependent parameters.  The $m_{\mu\mu}$ distribution is fit between \SI{2.8}{\giga\electronvolt} and \SI{70}{\giga\electronvolt} for all regions, except for CRb, which has a lower bound at \SI{15}{\giga\electronvolt}.  There is no contribution to the total background from the $\psi$ or $\Upsilon$ resonances.}
\scalebox{0.94}{
\begin{ruledtabular}
\begin{tabular}{lccc}
Parameter & $f_{\Upsilon}\left[\frac{\Upsilon}{\psi+\Upsilon}\right]$  $\left(\si[number-unit-product = {}]{\percent}\right)$ & $f_{\mathrm{Res}}\left[\frac{\psi+\Upsilon}{\mathrm{Total}}\right]$  $\left(\si[number-unit-product = {}]{\percent}\right)$ & $f_{t\bar{t}}\left[\frac{t\bar{t}}{\mathrm{Total}}\right]$ $\left(\si[number-unit-product = {}]{\percent}\right)$ \\\colrule
CRj & \num{32.6+-0.3} & \num{14.7+-0.1} & \num{6.1+-0.9} \\
CRb & N/A & N/A & \num{87.2+-5.1} \\
VR$\mu$ & \num{35.8+-6.0} & \num{18.8+-2.3} & \num{28.2+-3.2} \\
VR$e$ & \num{36.3+-9.2} & \num{12.2+-2.3} & \num{34.2+-3.6} \\
SR$\mu$ & \num{25.8+-4.9} & \num{15.2+-1.6} & \num{20.4+-4.1} \\
SR$e$ & \num{24.5+-6.6} & \num{11.8+-1.6} & \num{23.5+-5.0} \\
\end{tabular}
\end{ruledtabular}
}
\end{table}

\begin{figure}[tb]
\subfloat{\includegraphics[width=0.82\columnwidth]{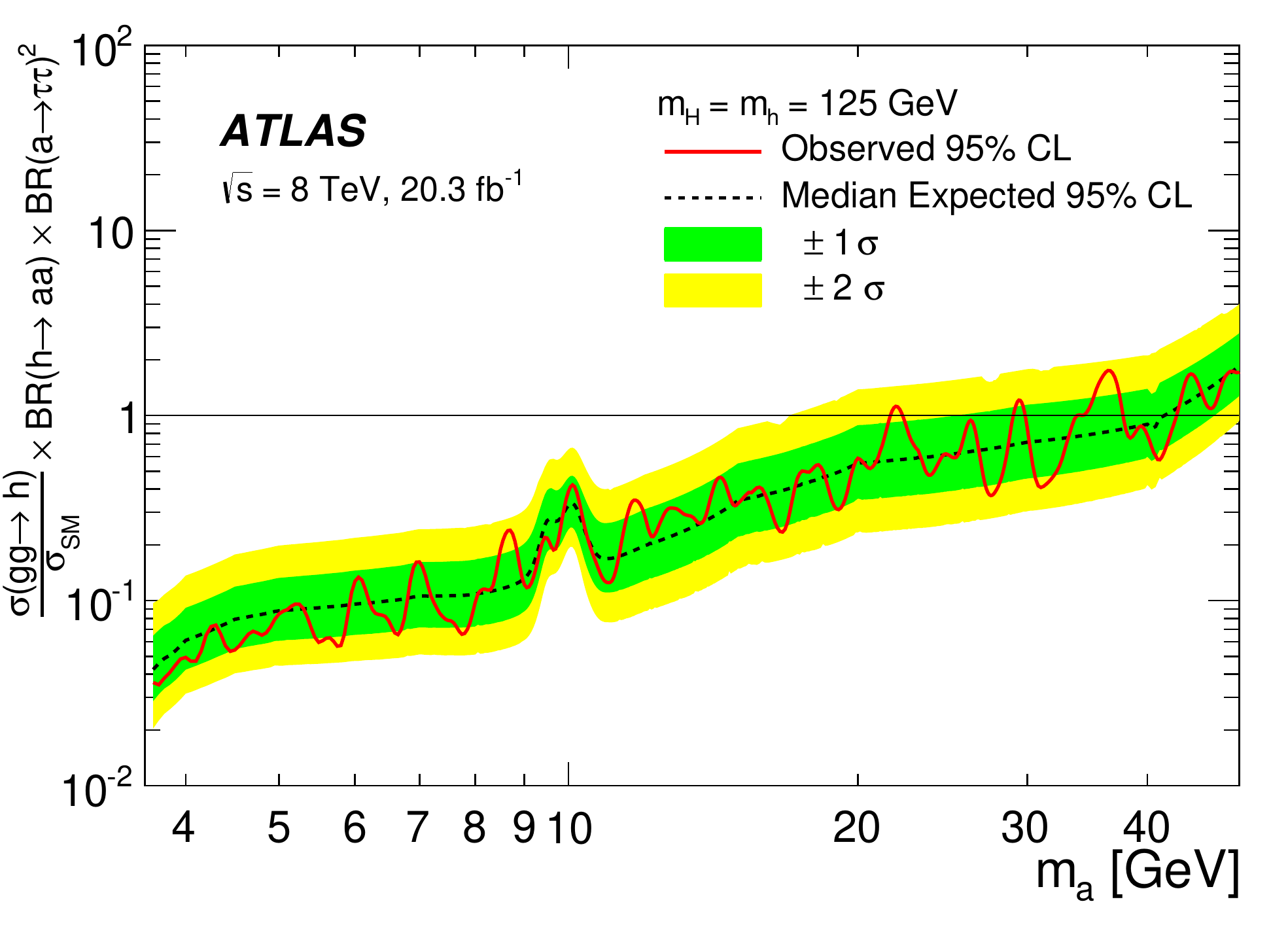}}\\
\subfloat{\includegraphics[width=0.82\columnwidth]{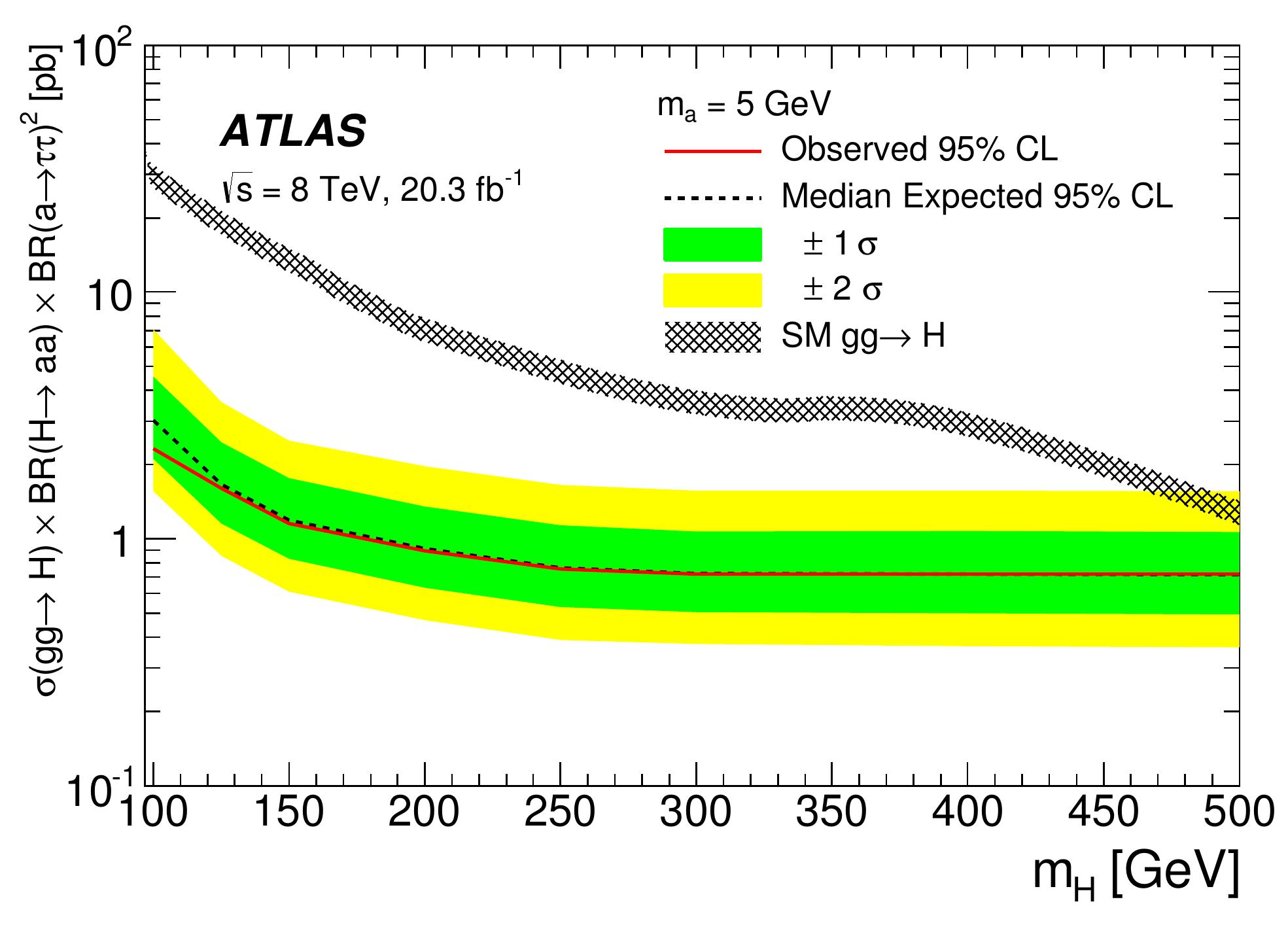}}
\caption[]{\small\label{f:lim}
Observed (solid red) and expected (dashed black) limits with the expected $\pm 1\sigma$ and $\pm 2\sigma$ bands shown in green and yellow respectively.  The top figure shows the expected and observed limits on the rate ($\sigma(gg\rightarrow{h})\times$BR$(h\rightarrow{aa})$) relative to the SM Higgs boson gluon-gluon fusion production cross section ($\sigma_{\mathrm{SM}}$) as a function of $m_a$ with $m_H$ set to \SI{125}{\giga\electronvolt}.  The limits are evaluated in \SI{50}{\mega\electronvolt} intervals below \SI{15}{\giga\electronvolt}, then \SI{100}{\mega\electronvolt} intervals up to \SI{30}{\giga\electronvolt}, and \SI{200}{\mega\electronvolt} intervals up to $m_a=$~\SI{50}{\giga\electronvolt}.  Shown in the bottom figure is the total rate ($\sigma(gg\rightarrow{H})\times$BR$(H\rightarrow{aa})$) as a function of $m_H$ with $m_a$ set to \SI{5}{\giga\electronvolt}, evaluated at \SI{50}{\giga\electronvolt} intervals from $m_{H}=$~\SIrange{100}{500}{\giga\electronvolt} and at $m_H=m_h=$~\SI{125}{\giga\electronvolt}.  The width of the black band in the bottom figure indicates the theoretical uncertainty on the SM $gg\rightarrow{H}$ cross section~\cite{CERNYellowReport3}.  In both figures, the observed and expected limits have been scaled by an $O(1)$ parameter, BR$(a\rightarrow\tau\tau)^2$, to account for the branching ratios assumed in this analysis and facilitate reinterpretation of the results.}
\end{figure}

  With no evidence to support the NMSSM hypothesis, a \SI[number-unit-product = {}]{95}{\percent} CL limit can be set using the $CL_s$ prescription~\cite{Read:2002hq}.  Figure~\ref{f:lim} shows the observed and expected limits on the rate ($\sigma(gg\rightarrow{h})\times$BR$(h\rightarrow{aa})$) relative to the SM Higgs boson gluon-gluon fusion production cross section ($\sigma_{\mathrm{SM}}$), calculated at NLO+NNLL precision~\cite{CERNYellowReport3}, as a function of $m_a$ with $m_H$ set to \SI{125}{\giga\electronvolt}.  The limits are evaluated in the same intervals used for the $p$-value scan.  Also shown in the figure is the total rate ($\sigma(gg\rightarrow{H})\times$BR$(H\rightarrow{aa})$) as a function of $m_H$with $m_a$ set to \SI{5}{\giga\electronvolt}, evaluated at \SI{50}{\giga\electronvolt} intervals from $m_{H}=$~\SIrange{100}{500}{\giga\electronvolt} and at $m_H=m_h=$~\SI{125}{\giga\electronvolt}.  In both panels of Fig.~\ref{f:lim}, the observed and expected limits have been scaled by BR$(a\rightarrow\tau\tau)^2$ to explicitly account for the branching ratios assumed in this analysis and facilitate reinterpretation of the results.

\FloatBarrier

\section{Conclusion}
\label{sec:conclusion}

A search for the decay of a scalar Higgs boson to two pseudoscalar $a$ Higgs bosons ($H\rightarrow{aa}$) in the context of the NMSSM is presented with LHC data corresponding to an integrated luminosity of \SI{20.3}{\per\femto\barn} of $pp$ collisions at $\sqrt{s}=$~\SI{8}{\tera\electronvolt}, collected in 2012 by the {ATLAS} experiment.  Final states are considered with two muons consistent with the decay of one $a$ boson as well as a third lepton ($e$ or $\mu$) and tracks, consistent with collimated $\tau$-leptons from the other $a$ boson.  A scan of the dimuon invariant mass distribution from \SIrange{3.7}{50}{\giga\electronvolt} shows no significant excess of data over SM backgrounds.  Limits are set assuming no coupling of the $a$ boson to quarks.  The observed \SI[number-unit-product = {}]{95}{\percent} CL upper limits on the production rate, $\sigma(gg\rightarrow{H})\times$BR$(H\rightarrow{aa})$, are consistent with the expected limit and are determined to be from \SIrange{2.33}{0.72}{\pico\barn}, for $m_{H}$ between \SI{100}{\giga\electronvolt} and \SI{500}{\giga\electronvolt} (and $m_a$ =~\SI{5}{\giga\electronvolt}).  A \SI[number-unit-product = {}]{95}{\percent} CL upper limit for the production of the $h$ boson and its decay rate to two pseudoscalar $a$ bosons is set for $m_a$ from \SIrange{3.7}{50}{\giga\electronvolt}, with the most stringent limit placed at \SI[number-unit-product = {}]{3.5}{\percent} for $m_a =$ \SI{3.75}{\giga\electronvolt}.

\begin{acknowledgements}


We thank CERN for the very successful operation of the LHC, as well as the
support staff from our institutions without whom ATLAS could not be
operated efficiently.

We acknowledge the support of ANPCyT, Argentina; YerPhI, Armenia; ARC,
Australia; BMWFW and FWF, Austria; ANAS, Azerbaijan; SSTC, Belarus; CNPq and FAPESP,
Brazil; NSERC, NRC and CFI, Canada; CERN; CONICYT, Chile; CAS, MOST and NSFC,
China; COLCIENCIAS, Colombia; MSMT CR, MPO CR and VSC CR, Czech Republic;
DNRF, DNSRC and Lundbeck Foundation, Denmark; EPLANET, ERC and NSRF, European Union;
IN2P3-CNRS, CEA-DSM/IRFU, France; GNSF, Georgia; BMBF, DFG, HGF, MPG and AvH
Foundation, Germany; GSRT and NSRF, Greece; RGC, Hong Kong SAR, China; ISF, MINERVA, GIF, I-CORE and Benoziyo Center, Israel; INFN, Italy; MEXT and JSPS, Japan; CNRST, Morocco; FOM and NWO, Netherlands; BRF and RCN, Norway; MNiSW and NCN, Poland; GRICES and FCT, Portugal; MNE/IFA, Romania; MES of Russia and NRC KI, Russian Federation; JINR; MSTD,
Serbia; MSSR, Slovakia; ARRS and MIZ\v{S}, Slovenia; DST/NRF, South Africa;
MINECO, Spain; SRC and Wallenberg Foundation, Sweden; SER, SNSF and Cantons of
Bern and Geneva, Switzerland; NSC, Taiwan; TAEK, Turkey; STFC, the Royal
Society and Leverhulme Trust, United Kingdom; DOE and NSF, United States of
America.

The crucial computing support from all WLCG partners is acknowledged
gratefully, in particular from CERN and the ATLAS Tier-1 facilities at
TRIUMF (Canada), NDGF (Denmark, Norway, Sweden), CC-IN2P3 (France),
KIT/GridKA (Germany), INFN-CNAF (Italy), NL-T1 (Netherlands), PIC (Spain),
ASGC (Taiwan), RAL (UK) and BNL (USA) and in the Tier-2 facilities
worldwide.

\end{acknowledgements}

\newpage
\begin{widetext}
\input{atlas_authlist}
\end{widetext}

\end{document}

%% file: atlas_authlist.tex
\begin{flushleft}
{\Large The ATLAS Collaboration}

\bigskip

G.~Aad$^{\rm 85}$,
B.~Abbott$^{\rm 113}$,
J.~Abdallah$^{\rm 151}$,
O.~Abdinov$^{\rm 11}$,
R.~Aben$^{\rm 107}$,
M.~Abolins$^{\rm 90}$,
O.S.~AbouZeid$^{\rm 158}$,
H.~Abramowicz$^{\rm 153}$,
H.~Abreu$^{\rm 152}$,
R.~Abreu$^{\rm 30}$
% ...
\bigskip
\\
% ...
$^{11}$ Institute of Physics, Azerbaijan Academy of Sciences, Baku, Azerbaijan\\
% ...
$^{30}$ CERN, Geneva, Switzerland\\
% ...
$^{85}$ CPPM, Aix-Marseille Universit{\'e} and CNRS/IN2P3, Marseille, France\\
% ...
$^{90}$ Department of Physics and Astronomy, Michigan State University, East Lansing MI, United States of America\\
% ...
$^{107}$ Nikhef National Institute for Subatomic Physics and University of Amsterdam, Amsterdam, Netherlands\\
% ...
$^{113}$ Homer L. Dodge Department of Physics and Astronomy, University of Oklahoma, Norman OK, United States of America\\
% ...
$^{151}$ Institute of Physics, Academia Sinica, Taipei, Taiwan\\
$^{152}$ Department of Physics, Technion: Israel Institute of Technology, Haifa, Israel\\
$^{153}$ Raymond and Beverly Sackler School of Physics and Astronomy, Tel Aviv University, Tel Aviv, Israel\\
% ...
$^{158}$ Department of Physics, University of Toronto, Toronto ON, Canada\\
% ...
\end{flushleft}
